\documentclass[runningheads]{llncs}

\usepackage[T1]{fontenc}
\usepackage{graphicx}
\usepackage{amsmath}
\usepackage{url}
\usepackage[table]{xcolor}
\usepackage{booktabs}
\usepackage{siunitx}
\usepackage[ruled,vlined,linesnumbered]{algorithm2e}
\SetKwComment{tcp}{//\ }{ }
\SetKwInput{KwIn}{Input}
\SetKwInput{KwOut}{Output}
\newcommand{\Call}[2]{\textnormal{#1}(#2)}
\usepackage{placeins}
\usepackage{enumitem}
\usepackage[colorlinks=true,citecolor=blue,linkcolor=blue,urlcolor=blue]{hyperref}
\begin{document}

\title{Similarity-Aware Client-Side Reordering for Server-Side Storage Compression in Distributed File Systems}
\titlerunning{Similarity-Aware Reordering for Server-Side Compression}

\author{Yuanhong Huang\inst{1,2} \and
Zicong Chen\inst{3}  \and Yili Ma\inst{2} \and Jinwu Yang\inst{2}
\and Yulong Chen \inst{2} \and Guangming Tan\inst{2}\and Dingwen Tao\inst{2}$^{(*)}$}

\authorrunning{Y. Huang et al.}
\institute{Hangzhou Institute for Advanced Study, University of Chinese Academy of Sciences, Hangzhou, China
\and
Institute of Computing Technology, Chinese Academy of Sciences, Beijing, China
\email{taodingwen@ict.ac.cn}
\and
School of Software, Beijing Jiaotong University, Beijing, China
}

\maketitle

\begin{abstract}
As growing data volumes strain the capacity of distributed file systems, server-side storage compression transparently reduces disk usage, but its effectiveness is limited when similar blocks land in separate compression records and cannot share matches. 
We present a client-side reordering
framework that transforms the physical block layout before persistence, so that
related blocks are co-located within the same compression scope without
replacing the existing storage-server compressor. It operates on
bounded reordering windows, periodically selects among candidate layouts
via lightweight compression of sampled data, and recovers the original logical order during
reads with compact window-local metadata. We implement the framework in the
SkyFS distributed file system, using the Zstandard (Zstd) compressor on the
OpenZFS storage backend. Results on diverse datasets show that the framework
reduces physical storage by up to 42.9\% relative to logical-order writes, while
achieving 94--108\% of baseline sequential-read throughput and 91--99\% of
random-read IOPS. Client-side layout construction is the primary additional
write cost, showing that adaptive similarity-aware reordering
substantially improves storage efficiency with acceptable overhead.
\keywords{Distributed file systems \and Server-side compression \and
Similarity-aware reordering \and Compression locality}
\end{abstract}

\section{Introduction}
\label{sec:introduction}

As global data volumes continue to grow, distributed file systems (DFSs) face
increasing capacity pressure \cite{statista2024worldwideData}.
Server-side storage compression addresses this pressure: it compresses data
before persistence, preserves standard file read/write interfaces, and directly
reuses mature compression engines. The storage system fully controls
compression policy, so upper-layer applications need not implement compression
logic or maintain persistent compressed formats.

Server-side storage compression commonly uses lossless techniques because persisted
data must be reconstructed exactly. Many compressors exploit repeated
byte sequences: LZ77 searches a bounded history, and Zstandard (Zstd) combines
matching with entropy coding~\cite{ziv1977lz77,collet2021zstd}. Their
effectiveness is constrained by compression locality. Since backends compress
records independently, similar blocks in different records cannot share
matches. Compression therefore depends on both content and physical
layout. Application write order usually follows logical structure or
production time, not similarity, leaving related blocks separated.

Deduplication, resemblance detection, and similarity detection identify
related content
\cite{broder1997resemblance,broder2000nearDuplicate,rabin1981fingerprinting,zhang2019finesse,zou2021odess,huang2024palantir,xu2026argus}.
Migratory Compression also shows that co-locating related data can improve
compressibility~\cite{lin2014migratory}. Prior work on such reorganization has focused mainly
on offline single-file processing or archival migration. An online
DFS write path must instead transform layout incrementally before persistence,
without waiting for complete files or replacing the compressor. It also cannot
defer a stream for a global layout or reorganize committed regions when the
layout changes.

We address these requirements with a similarity-aware client-side reordering
framework. Detectors provide similarity hints, while a common,
detector-independent generator constructs candidate layouts. The client uses
bounded windows and periodically sampled trial compression to select among
candidates. A reversible window-local map restores logical order for reads.
Applications retain the conventional file interface within the supported
workload scope, and the storage-server compressor remains unchanged. This separation lets detectors share layout construction,
selection, and recovery instead of embedding placement policy in each
detector.

We implement the framework in SkyFS~\cite{xing2009adaptive} with
OpenZFS~\cite{openzfs2024zfsprops} and Zstd, and evaluate it on archives,
virtual machine (VM) images, database files, and scientific data. Relative to NoReorder-128KB, it
reduces storage consumption by up to 42.9\% while retaining near-baseline
sequential and random-read performance. Client-side layout
construction is the main additional write-path cost.

The contributions of this paper are summarized as follows:
\begin{itemize}

\item We integrate bounded-window similarity-aware reordering with logical-order recovery into the online DFS data path, without replacing the compressor on the storage server.

\item We design a detector-independent, record-aware layout generator with
periodic sampled selection; detectors supply interchangeable hints, and the
common generator determines placement.

\item We implement the design in SkyFS/OpenZFS, evaluate physical-storage
savings, sequential I/O, random reads, and write-path overhead, and
achieve up to 42.9\% storage savings and near-baseline read performance.

\end{itemize}

\section{Background and Motivation}
\label{sec:background}

\subsection{Compression in Distributed File Systems}
\label{sec:compression}
Compression placement determines the resource being optimized.
Client-side transfer compression encodes data before transmission and decodes
upon arrival, primarily reducing network traffic. GlusterFS provides an
extensible client-side translator stack to support such transfer-time
transformations~\cite{redhat2026glusterfs}.

Server-side storage compression instead compresses ordinary file data at the
storage server or local backend before persistence. Lustre can use the ZFS file system to back its
object storage, while Ceph uses BlueStore as a local
backend~\cite{braam2002lustre,lustreWikiZfs,weil2006ceph,aghayev2019bluestore}.
This work targets server-side storage compression: the client changes block order without
compressing contents or reducing network bytes. The unchanged storage-server compressor
then processes the transformed block layout without further changes.

\subsection{Compression Locality and Similarity-Aware Reordering}
\label{sec:similarity-background}

\begin{figure}[ht]
  \centering
  \includegraphics[width=0.95\textwidth]{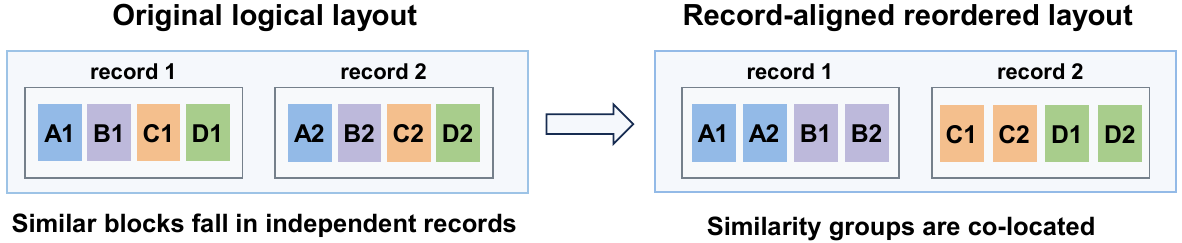}\\[-1mm]
  \caption{Compression-locality problem and opportunity. In the original
  layout, similar logical blocks occupy different compression records.
  Record-aligned reordering places them in a common compression scope, while a
  recovery map preserves logical order.}
  \label{fig:reordering-example}
\end{figure}

LZ77-based compressors~\cite{ziv1977lz77} improve compression on similar data
by encoding repeated byte sequences as compact (offset, length) references to
earlier matches in the bounded history. Zstd~\cite{collet2021zstd} combines this
matching with entropy coding, so more and longer matches generally produce
smaller compressed output. Since independently compressed records reset match
history, similar blocks in different records cannot reference each other;
co-locating them within the same record exposes their shared byte sequences to
the compressor and improves the compression ratio.
Figure~\ref{fig:reordering-example} illustrates this effect: record-aligned
reordering moves similar blocks into a shared compression scope, while a
recovery map preserves logical order.

MinHash and rolling fingerprints provide the basis for estimating such
overlap~\cite{broder1997resemblance,broder2000nearDuplicate,rabin1981fingerprinting}.
Odess computes rolling hashes over overlapping windows, retains a
content-defined sample, applies several independent linear transformations,
and takes the minimum under each as a raw feature; groups of raw features are
hashed into super-features~\cite{zou2021odess}. Palantir uses the same
N-transform raw features in several grouping
tiers---$(k,s)=(3,4),(4,3),(6,2)$ by default---forming $k$ super-features from
groups of $s$ features, and queries from stricter to looser tiers to broaden
coverage while prioritizing higher-similarity
matches~\cite{huang2024palantir}.

Argus instead content-samples fine-grained Gear hashes, partitions the sampled
values into disjoint content-dependent bins to suppress uninformative features,
applies a linear transformation within each bin, and selects the transformed
minimum as that bin's plain feature, without combining multiple plain features
into a super-feature~\cite{xu2026argus}.

These methods provide alternative position-specific descriptors for a block.
Section~\ref{sec:record-aware-layout} uses matching descriptors as
hash-lookup hints for forming similarity groups within each reordering window.

\subsection{Challenges of Online Similarity-Aware Reordering}
\label{sec:motivation}

\textbf{Bounded online processing.} The client cannot analyze an entire large
or appended file before writing. Reordering must use bounded units and handle
partial units produced by \texttt{flush} or \texttt{close}.

\textbf{Adaptive layout selection.} Different detectors and data phases favor
different layouts. A fixed choice cannot adapt, but exhaustive per-unit
evaluation is too expensive, motivating lightweight periodic selection.

\textbf{Logical-order recovery.} Reordering breaks the logical-to-physical
mapping. Reads require bounded recovery metadata without a large file-wide index
or excessive read amplification. Persisted metadata must remain sufficient to
interpret the transformed layout after writes complete.

Increasing the backend record size expands the compression scope uniformly,
whereas reordering selectively co-locates related data. Larger records also
increase the decompression unit for fine-grained reads, making them an
important baseline rather than an equivalent replacement.

These challenges motivate bounded-window reordering, periodic sampled layout
selection, and compact window-local recovery.

\section{Design and Implementation}
\label{sec:design}

\subsection{System Overview}
\label{sec:overview}

\begin{figure}[t]
  \centering
  \includegraphics[width=\textwidth]{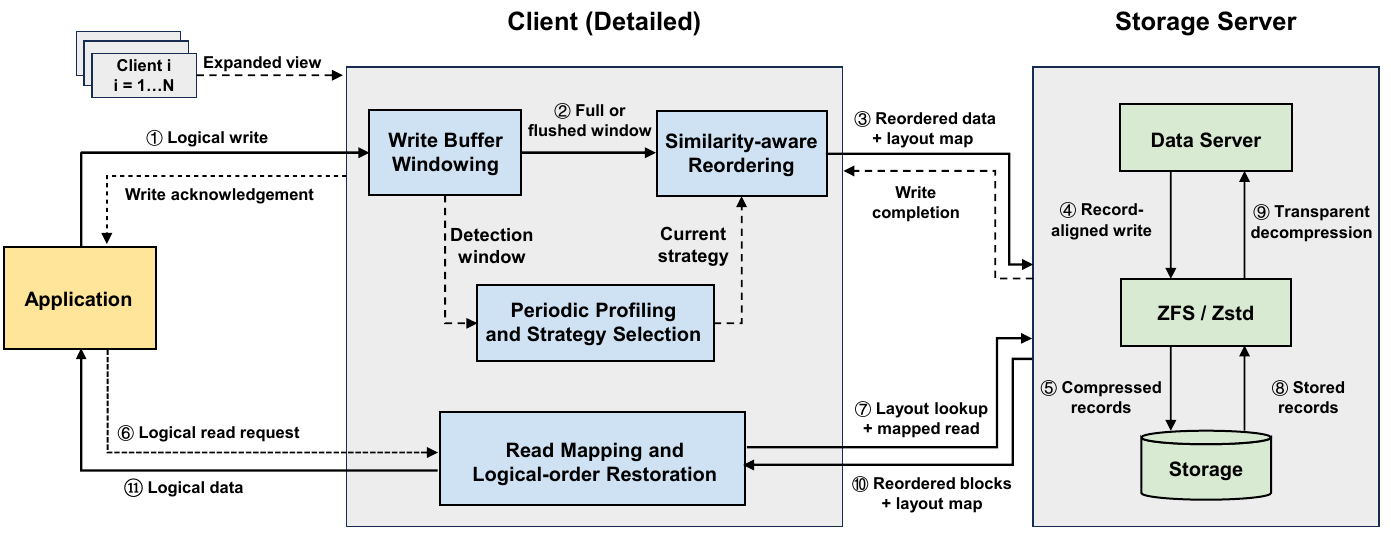}\\[-1mm]
  \caption{System architecture and numbered write/read paths. Solid arrows
  denote application data, and dashed arrows denote selection and completion
  metadata.}
  \label{fig:overview}
\end{figure}

Figure~\ref{fig:overview} shows the end-to-end data path, connecting the
client buffer, similarity detectors, layout generator, selector, and mapping
module to the unchanged server-side storage compression path.

\textbf{Write path.} The write path proceeds in three stages. First, logical
writes accumulate in 64~MB reordering windows. Once a window is complete---or
emitted early due to a \texttt{flush}, \texttt{close}, or timeout---the client
applies the current candidate layout, grouping similar blocks and packing them
into 128~KB records. Finally, the client writes the reordered data and its
recovery map through the existing file interface; OpenZFS forms and compresses
the backend records and completes the write normally.

\textbf{Read path.} Reads reverse this process in two stages. A logical request
is first split at reordering-window boundaries when necessary. The mapping
module then loads each window's recovery map, translates logical offsets to
physical positions, and issues mapped reads. OpenZFS decompresses the
containing records, and the client places the returned blocks at their
application offsets.

The prototype targets large-file sequential-ingestion and append workloads;
arbitrary in-place overwrites and concurrent same-file modifications are
outside its scope. Within this scope, applications observe the conventional
file interface and unchanged file contents.

\subsection{Within-Window Similarity Grouping and Record Layout}
\label{sec:record-aware-layout}

The layout generator processes blocks in file-offset order using a greedy
representative policy where the first block in a group serves as its base
block (\texttt{BaseChunkID}). All three detectors follow a common grouping
procedure: for each incoming block, the generator looks up its features in
hash tables, adding matches to an existing group and starting a new group
on a miss. Odess and Argus use three feature tables (super-features and
plain features, respectively); Palantir cascades through three tiered tables.

\begin{figure}[b]
  \centering
  \includegraphics[width=\textwidth]{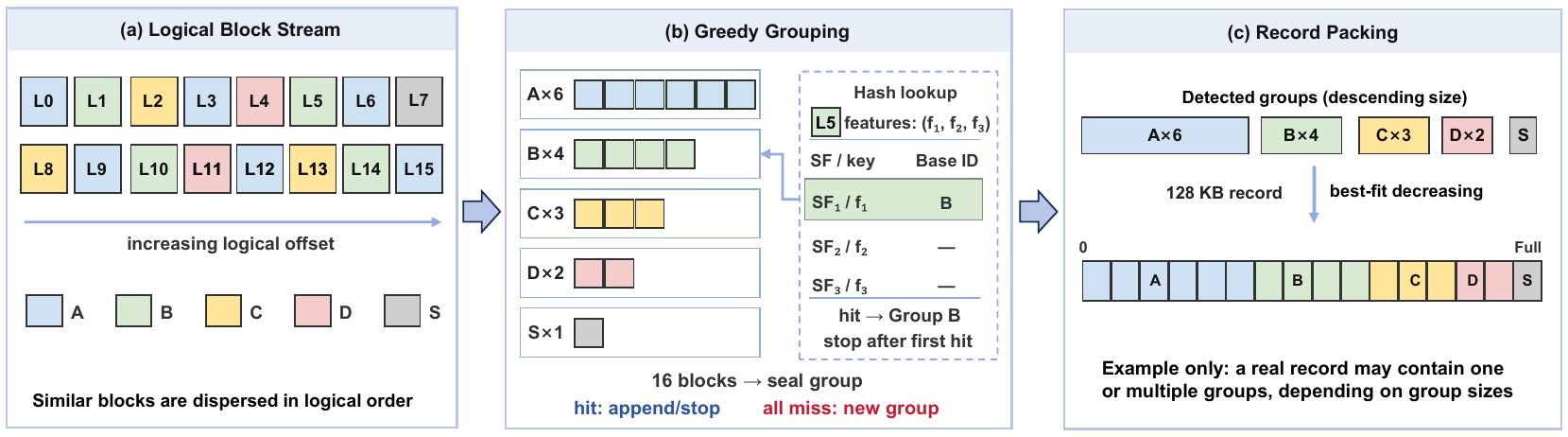}\\[-1mm]
  \caption{Within-window similarity grouping and record packing.
  This example shows the case where no group
  reaches the sealing threshold; when a group reaches 16~blocks, it is sealed
  and directly forms a full record.}
  \label{fig:grouping}
\end{figure}

Figure~\ref{fig:grouping} illustrates this procedure. A 128~KB compression
record holds sixteen 8~KB blocks. When a group reaches this size, it is sealed
and the index entries of its base block are removed so subsequent blocks cannot
join it.

Sealed full groups directly form compression records. The client packs the
remaining multi-block groups intact using best-fit decreasing, then fills
residual positions with unmatched singletons in logical order. If intact groups
exceed the minimum record count implied by their combined size---indicating
that fragmentation would waste record capacity---the smallest group is
dissolved and packing is retried. The procedure adds no padding, preserves
logical order within retained groups, and permits only the final record to be
partial.

\subsection{Periodic Layout Selection}
\label{sec:periodic-selection}

A fixed detector cannot adapt when workload patterns change across files or data
phases. The client maintains
an incumbent layout, starting with NoReorder-128KB, and re-evaluates the full set of
candidates
every $T=16$ complete reordering windows using a switching threshold
$\epsilon=0.05$. We fix $T$, $\epsilon$, and the sampling configuration below
across all experiments; automatic tuning is left to future work.

\vspace{0.5ex}
\begin{algorithm}[H]
\caption{Periodic Similarity-Aware Layout Selection}
\label{alg:periodic-selection}
\small
\KwIn{Window $w_k$; index $k$; incumbent $c$; candidates $\mathcal{S}$;
  period $T$; threshold $\epsilon$}
\KwOut{Layout $s^\star$; block order $L$; recovery map $P$}
Initialize $s^\star \gets c$\;
\If{$(k-1)\bmod T = 0$}{
  $X_1 \gets \Call{StratifiedSample}{w_k,8}$ \tcp{4~MB pilot}
  \For{$s \in \mathcal{S}$}{
    $C_s^{(1)} \gets \Call{TrialCompress}{s, X_1}$\;
  }
  $b \gets \arg\min_{s \in \mathcal{S}} C_s^{(1)}$\;
  \If{$C_b^{(1)} < (1-\epsilon)C_c^{(1)}$}{
    $X_2 \gets \Call{StratifiedSample}{w_k \setminus X_1,24}$\tcp{12~MB verification}
    \For{$s \in \{b,c\}$}{
      $C_s^{(2)} \gets \Call{TrialCompress}{s, X_2}$\;
    }
    \If{$C_b^{(1)}+C_b^{(2)} < (1-\epsilon)(C_c^{(1)}+C_c^{(2)})$}{
      $s^\star \gets b$\;
    }
  }
}
$L \gets \Call{FullLayout}{s^\star, w_k}$\;
$P \gets \Call{PhysicalToLogicalMap}{L}$\;
\Return{$(s^\star, L, P)$}\;
\end{algorithm}
\vspace{0.5ex}

Selection uses a two-stage stratified sampling procedure within a single
reordering window, summarized in Algorithm~\ref{alg:periodic-selection}. The
window is partitioned into 64 strata of 1~MB each. In the pilot stage, the
client draws eight 8~KB blocks per stratum---4~MB in total---and
trial-compresses every candidate layout with Zstd level~3. The best-performing
candidate becomes the challenger only if it exceeds the incumbent by at least
$\epsilon$; otherwise the incumbent is retained without further evaluation.
When a challenger emerges, the verification stage draws an additional 24
disjoint blocks per stratum (12~MB total) and evaluates only the challenger
against the incumbent. The client switches to the challenger only when both
stages jointly favor it by the same margin. Across four candidates, sampling
totals at most 40~MB per 1~GB selection interval ($\approx$4\%), counting bytes
supplied to trial compression. Non-selection and partial windows reuse
the incumbent without evaluation; partial windows also leave the selection
counter untouched until the next selection interval.

\FloatBarrier
\subsection{Window-Local Reversible Layout}
\label{sec:reversible-layout}

Reordering improves compression but breaks the direct correspondence between
logical and physical block positions. Each window therefore carries its own
recovery metadata. A complete 64~MB window contains 8192 8~KB logical blocks.
We persist a physical-to-logical (P2L) map---one logical identifier per
physical block position---because sequential recovery, the common read pattern,
consumes P2L directly and places each returned block at its application offset.
A two-byte identifier yields an unencoded 16~KB map, approximately 0.025\% of
the window. Unchanged runs are interval-encoded, and the map is further
compressed by OpenZFS.

Random reads construct the inverse logical-to-physical (L2P) map $L2P[P(i)]=i$ in memory on demand; the
client may cache or evict this inverse. The P2L map is persisted as an ordinary
sidecar object through the existing storage path, and file metadata records its
layout generation and index root so a client can locate the relevant recovery
map. For an 8~KB random read, the client translates the logical offset and
requests the corresponding physical block; OpenZFS decompresses the containing
backend compression record.

Crash consistency reuses the existing SkyFS commit mechanism. For each emitted
window, the client first writes the reordered data and its P2L map as immutable
objects. After both writes complete, a single metadata update publishes their
locations and layout generation. Readers access only windows referenced by
committed metadata; objects produced by an interrupted write remain unreferenced
and can be reclaimed. As a result, the reordering layer adds no cross-server
coordination beyond what SkyFS already provides.

\section{Evaluation}
\label{sec:evaluation}

We evaluate storage savings (Section~\ref{sec:storage-saving}),
periodic-selection effectiveness (Section~\ref{sec:selection-eval}), and
end-to-end overheads including sequential I/O, random-read IOPS, and
write-path breakdown (Section~\ref{sec:overhead}).

\subsection{Experimental Setup}
\label{sec:setup}

\textbf{Platforms.}
Our SkyFS prototype runs on two clients (2× AMD EPYC~9354, 512~GB DRAM, Ubuntu~22.04.5), one metadata server, and four storage nodes (2× Intel Xeon~4314, 128~GB DRAM, 12× 7~TB NVMe SSD, Ubuntu~22.10), connected by 64~Gb/s InfiniBand/RDMA. The prototype is implemented in C++17 with OpenZFS~2.1.7 \cite{openzfs2024zfsprops}.

\textbf{Baselines.}
NoReorder-128KB preserves logical order with 128~KB records. The Odess-based, Palantir-based, and Argus-based layouts each use a fixed detector;
Proposed periodically selects among them and NoReorder-128KB, using 128~KB
records. All configurations compress with Zstd at level~3.

NoReorder-512KB and NoReorder-1MB expand the backend compression scope uniformly
without client-side reordering; we include them in the storage-saving and
random-read experiments to quantify whether larger records alone can match
selective co-location.

\textbf{Datasets.}
Table~\ref{tab:datasets} summarizes four datasets spanning logs and source code,
VM images, database tablespaces, and scientific data. We retain native byte
order and archive TAR members deterministically. VM is used for
periodic-selection analysis.

\textbf{Methodology.}
Each space experiment uses a fresh ZFS dataset with the specified record size.
Compression ratio is raw logical size divided by the combined physical
footprint of compressed data and persistent recovery metadata.
All performance results are the means of five repetitions, and error bars
denote one standard deviation.

\begin{table}[t]
\caption{Evaluation datasets.}
\label{tab:datasets}
\centering
\small
\begin{tabular}{p{0.13\textwidth}p{0.13\textwidth}p{0.72\textwidth}}
\hline
Name & Size & Description \\
\hline
TAR & 5.36~GB & Linux systemd journal logs \cite{poettering2012journal}
and GCC, GDB, and Emacs source code from the GNU archives \cite{gnuArchives},
packaged as a TAR archive. \\
VM & 28.01~GB & Eight virtual-machine images, including CentOS, Fedora, and
Debian, obtained from OSBoxes \cite{osboxesVMwareImages}. \\
IBD & 24.50~GB & MySQL InnoDB \texttt{.ibd} files with database compression
disabled\cite{huang2026ibd}. \\
EXAFEL & 1.11~GB & Light-source data from the SDRBench scientific data
suite \cite{zhao2020sdrbench}. \\
\hline
\end{tabular}
\end{table}

\subsection{Storage Savings}
\label{sec:storage-saving}

\begin{table}[t]
  \caption{Compression ratio in the storage-saving experiment. The
  denominator is the combined physical footprint of file data and persistent
  recovery maps.}
  \label{tab:storage-saving}
  \centering
  \small
  \renewcommand{\arraystretch}{1.12}
  \setlength{\tabcolsep}{9pt}
  \begin{tabular}{@{}l *{4}{S[table-format=1.2]}@{}}
    \toprule
    & \multicolumn{4}{c}{Compression Ratio} \\
    \cmidrule(lr){2-5}
    Method & {TAR} & {VM} & {IBD} & {EXAFEL} \\
    \midrule
    NoReorder-128KB        & 4.31 & 2.27 & 3.91 & 1.13 \\
    NoReorder-512KB   & 4.59 & 2.35 & 3.94 & 1.13 \\
    NoReorder-1MB     & 4.71 & 2.39 & 4.01 & 1.13 \\
    Odess-based Layout    & 4.89 & 3.10 & 6.43 & 1.13 \\
    Palantir-based Layout & 4.90 & 3.18 & 6.29 & 1.13 \\
    Argus-based Layout    & 5.06 & 3.25 & 4.10 & 1.13 \\
    \rowcolor{gray!10}
    \textbf{Proposed} &\textbf{5.16} & \textbf{3.49} & \textbf{6.85} & \textbf{1.14} \\
    \bottomrule
  \end{tabular}
\end{table}

Table~\ref{tab:storage-saving} reports the compression ratio for each
configuration. Proposed achieves the highest ratio on all four
datasets, reducing net physical storage relative to NoReorder-128KB by 42.9\% on
IBD, 35.0\% on VM, 16.5\% on TAR, and 0.9\% on EXAFEL, whose already-encoded
scientific measurements leave little inter-block redundancy to exploit.

Larger records improve NoReorder modestly---NoReorder-1MB reaches 4.01$\times$
on IBD versus 6.85$\times$ for Proposed, a 70.8\% gap because high-similarity
database blocks are too dispersed for uniform expansion to capture. On VM and
TAR, Proposed achieves 46.0\% and 9.6\% higher ratios than NoReorder-1MB,
respectively. Selective co-location thus provides larger gains than uniformly
increasing the record size on compressible workloads.

No single fixed detector dominates across datasets: Argus leads on TAR and VM,
Odess leads on IBD, and all tie on EXAFEL. Proposed surpasses the best fixed-layout compression
ratio by 0.9--7.4\%, confirming that periodic selection adapts to
dataset-specific detector strengths.

\subsection{Effectiveness of Periodic Selection}
\label{sec:selection-eval}

\begin{figure}[t]
  \centering
  \begin{minipage}[b]{0.28\textwidth}
    \centering
    \includegraphics[width=\linewidth]{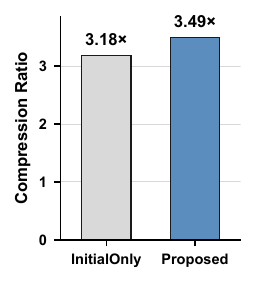}\\[-1mm]
    \textbf{(a)}
  \end{minipage}\hfill
  \begin{minipage}[b]{0.72\textwidth}
    \centering
    \includegraphics[width=\linewidth]{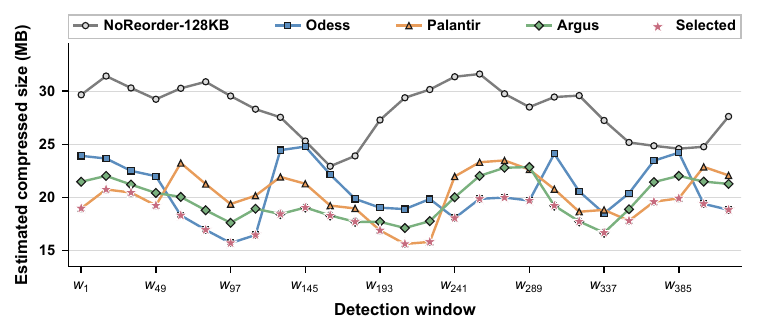}\\[-1mm]
    \textbf{(b)}
  \end{minipage}
  \caption{Periodic-selection effectiveness on the VM dataset.
  (a) Complete-workload net compression ratio of InitialOnly and Proposed.
  (b) Pilot trial-compressed sizes at reordering windows; stars mark the
  post-evaluation selected candidates.}
  \label{fig:periodic-selection}
\end{figure}

To isolate the benefit of periodic reselection, we compare Proposed (which
re-evaluates every $T=16$ complete windows) against InitialOnly, which selects
once at $w_1$ and reuses that decision for all subsequent windows. Both start
from NoReorder-128KB and use the same two-stage selector.

Figure~\ref{fig:periodic-selection}a shows the result: on VM,
Proposed achieves 3.49$\times$ versus 3.18$\times$ for InitialOnly, an 8.9\%
reduction in physical storage.
Figure~\ref{fig:periodic-selection}b reveals why: the pilot-preferred candidate
shifts among Palantir, Odess, and Argus as the workload progresses, reflecting
changing data characteristics within the VM image (stars mark verified
candidates retained after the 5\% threshold). A single detector chosen at file
start is thus suboptimal; periodic reselection exploits phase changes by
capturing complementary detector strengths across data phases. The other
datasets exhibit similar detector shifts across windows, suggesting that
periodic reselection generalizes beyond VM images.

\subsection{End-to-End Overhead}
\label{sec:overhead}

\textbf{Sequential I/O.}
We measure sequential throughput for all configurations at 2--64 client
threads, reported in uncompressed application GB/s.

\begin{figure}[t]
  \centering
  \includegraphics[width=\textwidth]{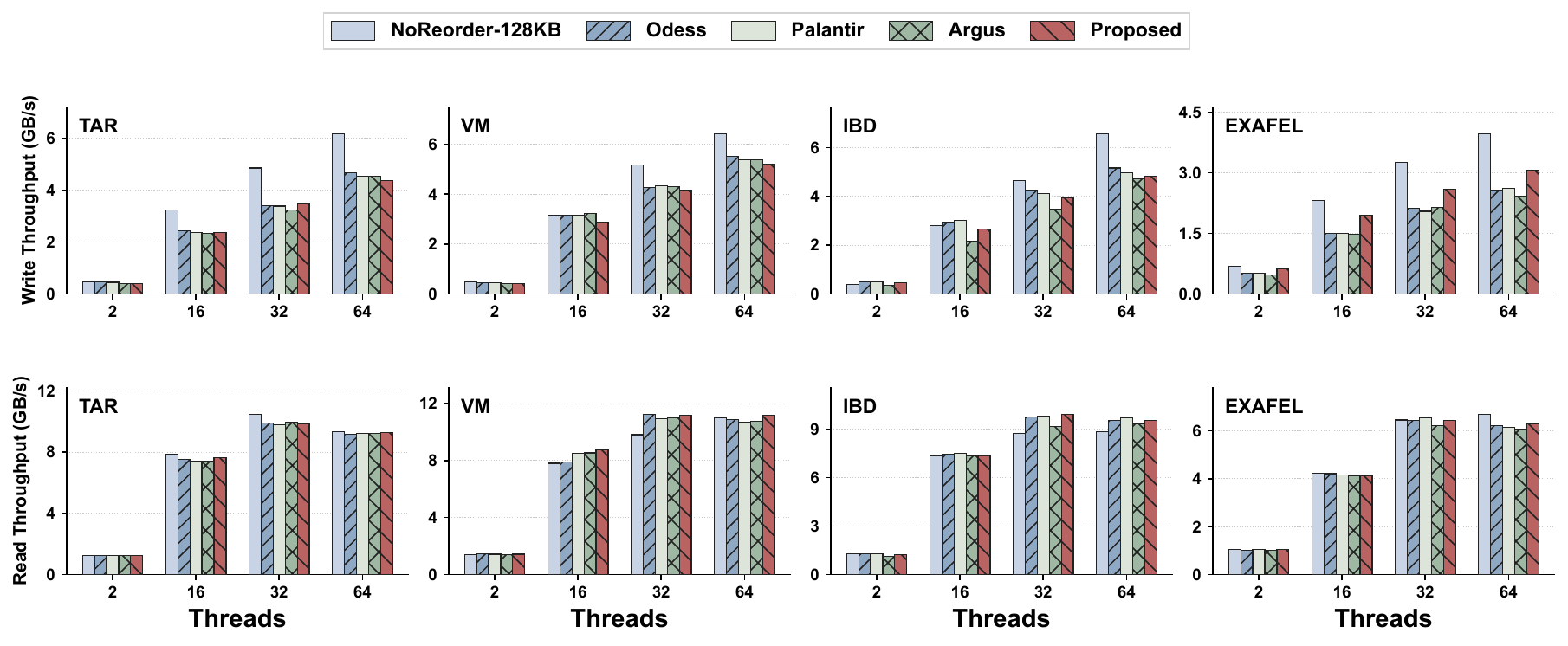}\\[-2mm]
  \caption{End-to-end sequential throughput. Top: writes; bottom: reads.}
  \label{fig:sequential-throughput}
\end{figure}

Figure~\ref{fig:sequential-throughput} presents write (top) and read (bottom)
throughput across all four datasets. On the write path, Proposed achieves
70.8--81.1\% of NoReorder-128KB's throughput at 64 threads; layout construction
accounts for the majority of this gap. IBD at two threads is the only
configuration where Proposed slightly outperforms NoReorder-128KB (0.45 versus
0.41~GB/s), suggesting that reduced downstream compression cost can offset
reordering overhead at low concurrency.

Read throughput is substantially closer to baseline, since detectors do not
execute on the read path and all layouts share the same logical-order recovery.
At 64 threads, Proposed achieves 94.4--107.5\% of NoReorder-128KB's read throughput;
across all thread counts and datasets, the largest decrease is 5.6\%.

\textbf{Random-read IOPS.}
We evaluate 8~KB random-read IOPS for NoReorder-128KB, NoReorder-512KB,
NoReorder-1MB, and Proposed at 2--64 threads, with ZFS data cache cold and
client-side L2P maps pre-warmed. Offsets are drawn uniformly; each thread
issues one synchronous request. Before each measurement, the ZFS ARC (Adaptive Replacement Cache) is dropped,
the client loads layout metadata and constructs inverse L2P maps, and 10,000
untimed random requests warm the L2P maps. Client data caching is bypassed;
results reflect DFS-path IOPS under cold data cache.

\begin{figure}[t]
  \centering
  \includegraphics[width=\textwidth]{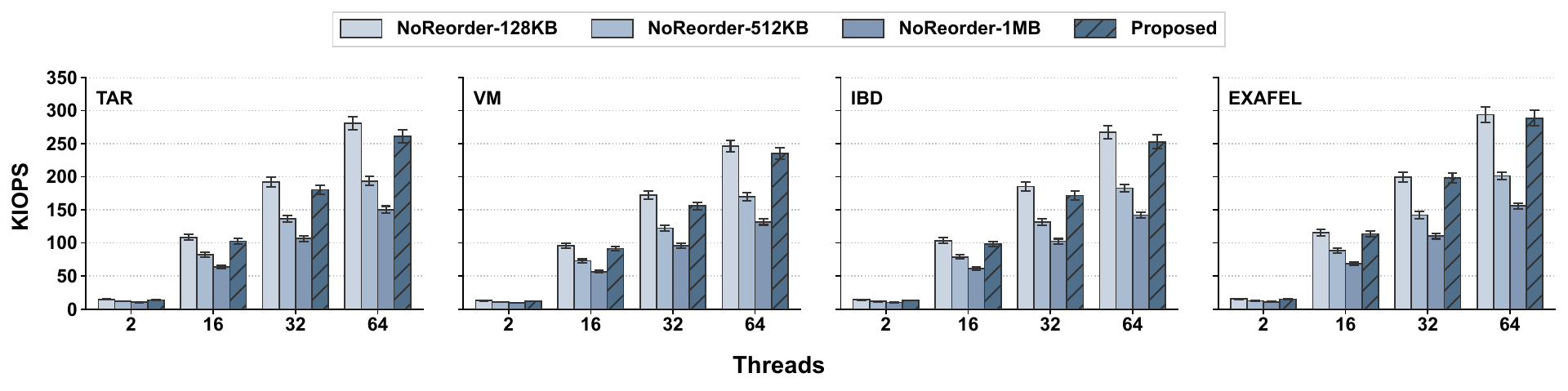}\\[-2mm]
  \caption{8~KB random-read IOPS for NoReorder-128KB, NoReorder-512KB,
  NoReorder-1MB, and Proposed.}
  \label{fig:random-read-iops}
\end{figure}

As shown in Figure~\ref{fig:random-read-iops}, at 64 threads Proposed sustains
235--289~KIOPS across the four datasets, retaining 92.9--98.2\% of
NoReorder-128KB (lowest on TAR, highest on EXAFEL). Across all thread counts
and datasets, retention ranges from 90.7\% to 99.4\%.

Larger records incur a progressively higher random-read penalty under cold data
cache. At 64 threads, NoReorder-512KB retains 68.4--69.0\% of
NoReorder-128KB's IOPS, and NoReorder-1MB retains only 53.1--53.5\%, because
each 8~KB request triggers decompression of a 512~KB or 1~MB record. In
contrast, Proposed avoids this decompression amplification through its
128~KB record granularity, and its IOPS exceeds NoReorder-512KB by 35--43\%
and NoReorder-1MB by 74--85\%. The measured overhead captures steady-state L2P
lookup, record decompression, and logical-order recovery, excluding one-time
metadata loading and warm-up.

\textbf{Write-path breakdown.}
To quantify the contribution of individual components, we instrument
non-overlapping phases of a VM write at 64 threads. Four phases sum to the
total completion time: Reorder , Network, Compression , and Software overhead.
Reorder is zero for NoReorder-128KB.

\begin{figure}[t]
  \centering
  \includegraphics[width=0.96\textwidth]{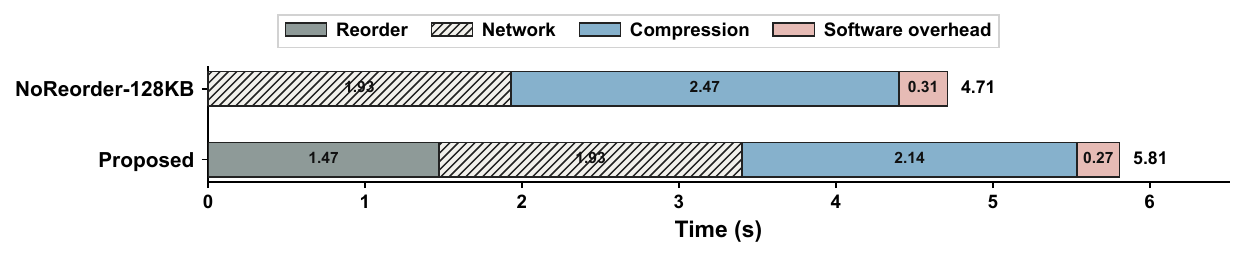}\\[-2mm]
  \caption{Write-path time breakdown on VM at 64 threads.}
  \label{fig:write-breakdown}
\end{figure}

As shown in Figure~\ref{fig:write-breakdown}, NoReorder-128KB completes in 4.71~s
and Proposed in 5.81~s. Network time is unchanged at 1.93~s.
Compression time decreases from 2.47~s to 2.14~s, reflecting the improved
compressibility of the reordered layout. Software overhead changes marginally
from 0.31~s to 0.27~s. Reorder adds 1.47~s, accounting for 25.3\% of
Proposed's total time and making overall completion 23.3\% longer than
NoReorder-128KB. Layout construction is therefore the dominant additional cost on the
write path.

\section{Conclusion and Future Work}
\label{sec:conclusion}

We presented a detector-independent client-side reordering framework that
co-locates similar blocks within shared compression scopes. Since the
storage-server compressor is untouched, the framework deploys without
server modifications. Across four datasets, it reduces physical storage by
up to 42.9\%; periodic selection saves a further 8.9\% on VM over one-shot
layout selection (InitialOnly). At 64 threads, write throughput retains
70.8--81.1\% of NoReorder-128KB, a gap that narrows---and can even
reverse---at low concurrency, while read throughput and random-read IOPS
remain at 92.9--98.2\% of baseline. The gains are workload-dependent:
reordering works best for large files with substantial inter-block
similarity---database tablespaces and VM images---while already-encoded
data such as EXAFEL benefits little. The dominant remaining cost is
client-side layout construction; future work includes faster detection and
grouping, small-file and mixed-workload support, and in-place overwrites of
committed windows.  

\textbf{Acknowledgments.} This work was supported by the National Natural Science Foundation of China through the Excellent Young Scientists Fund Program (Overseas) (Project No. E401470), the National Key Research and Development Program of China (Grant Nos. 2020YFA0907000 and 2025YFB30037002), the National Natural Science Foundation of China (Grant Nos. 62032023 and T2125013).

\textbf{Disclosure of Interests.} The authors have no competing interests to declare that are relevant to the content of this article.

\bibliographystyle{splncs04}
\bibliography{reference}

\end{document}